\documentclass[reprint,pra,aps,showpacs,amssymb,amsmath]{revtex4-1}
\usepackage{graphics}
\usepackage{bm}
\usepackage{graphicx}
\usepackage{amsmath}
\usepackage{amssymb}
\usepackage{epsfig}
\begin{document}
\title{Dynamics of entanglement between two harmonic modes in stable and unstable regimes}
\author{L. Reb\'on, N. Canosa, R. Rossignoli}
\affiliation{Departamento de F\'{\i}sica-IFLP, Universidad Nacional de La Plata,
C.C.~67, 1900 La Plata, Argentina}
\begin{abstract}

The exact dynamics of the entanglement between two harmonic modes generated by
an angular momentum coupling is examined. Such system arises when considering a
particle in a rotating anisotropic harmonic trap or a charged particle in a
fixed harmonic potential in a magnetic field, and exhibits a rich dynamical
structure, with stable, unstable and critical regimes according to the values
of the rotational frequency or field and trap parameters. Consequently, it is
shown that the entanglement generated from an initially separable gaussian
state can exhibit quite distinct evolutions, ranging from quasiperiodic
behavior in stable sectors to different types of unbounded increase in critical
and unstable  regions. The latter lead respectively to a logarithmic and linear
growth  of the entanglement entropy with time. It is also shown that
entanglement can be controlled by tuning the frequency, such that it can be
increased, kept constant or returned to a vanishing value just with stepwise
frequency variations. Exact asymptotic expressions for the entanglement entropy
in the different dynamical regimes are provided.
\end{abstract}
\maketitle

\section{Introduction}
The investigation of entanglement dynamics and growth in different physical
systems is of great current interest \cite{SLRD.13,BPM.12,DPSZ.12}. Quantum
entanglement is well known to be an essential resource for quantum
teleportation \cite{Be.93} and pure state based quantum computation
\cite{NC.00}, where its increase with system size is necessary to achieve an
exponential speedup over classical computation \cite{JL.03,Vi.03}. And a large
entanglement growth with time after starting from a separable state indicates
that the system dynamics cannot be simulated efficiently by classical means
\cite{SWVC.08}, turning it suitable for quantum simulations.

The aim of this work is to examine the dynamics of the entanglement between two
harmonic modes generated by an angular momentum coupling, and its ability to
reproduce typical regimes of entanglement growth in more complex many body
systems, when starting from an initial separable gaussian state. The latter can
be chosen, for instance, as the ground state of the non-interacting part of the
Hamiltonian, thus reproducing the typical quantum quench scenario
\cite{SLRD.13,BPM.12,SWVC.08}. The present system can be physically realized by
means of a charged particle in a uniform magnetic field within a harmonic
potential or  by a particle confined in a rotating harmonic trap
\cite{Va.56,FK.70,RS.80,BR.86},  where the field or rotational frequency
provides an easily controllable coupling strength. Accordingly, it has been
widely used in quite different physical contexts, such as rotating nuclei
\cite{RS.80,BR.86}, quantum dots in a magnetic field \cite{MC.94} and fast
rotating Bose-Einstein condensates within the lowest Landau level approximation
\cite{LNF.01,OO.04,AF.07,ABL.09,ABL.09,ABD.05,BDS.08}. In spite of its
simplicity, the model is able to exhibit a rich dynamical structure
\cite{RK.09}, with both stable and distinct types of unstable regimes,
characterized by bounded as well as unbounded dynamics, when considering all
possible values of the field or frequency in a general anisotropic potential.
Nonetheless, being a quadratic Hamiltonian in the pertinent coordinates and
momenta, the dynamics can be determined analytically in all regimes, and the
entanglement between modes can be evaluated exactly through the gaussian state
formalism \cite{WW.01,AEPW.02,ASI.04,BvL.05,WPPCRSL.12}. For the same reason,
the Hamiltonian is also suitable for simulation with optical techniques
\cite{PE.94}.

The main result we will show here is that due its non-trivial dynamical
properties, the entanglement dynamics in the previous model can exhibit
distinct regimes, including a quasiperiodic evolution in dynamically stable
sectors, different types of logarithmic growth at the border between stable and
unstable sectors (critical regime) and  a linear increase in dynamically
unstable sectors. The model is then able to mimic the three typical regimes for
the entanglement growth with time after a quantum quench, arising in spin $1/2$
chains with Ising type couplings, according to the results of refs.\
\cite{SLRD.13,SWVC.08},  which show a linear growth for short range couplings,
a logarithmic growth for long range interactions and an oscillatory behavior
for nearly infinite range interactions, when considering a half-chain
bipartition. We also mention that the static ground state entanglement of
the present model also exhibits critical behavior at the border of instability
\cite{LR.11}.  Mode entanglement dynamics in related harmonic models within
stable regimes were previously studied in \cite{HMM.03,NL.05,CN.08}, while
critical behavior and entanglement in ultrastrong-coupled oscillators
(through a different interaction) were considered  in \cite{SGLK.12}. Other
relevant aspects of entanglement dynamics and generation in spin systems were
discussed in \cite{RS.07,AOPFP.04,HK.05,AFOV.08}.

In  sec.\ \ref{II} we discuss the exact dynamics of the system and  describe
the different regimes arising for strong coupling in anisotropic potentials.
The entanglement evolution in gaussian states is then examined in detail in
sec.\ \ref{III}, including its exact evaluation through the covariance matrix
formalism and the exact asymptotic behavior in the distinct dynamical regimes.
Explicit results,  including the possibility of entanglement control through a
stepwise varying frequency, are also shown. Conclusions are finally provided in
\ref{IV}.

\section{Model and exact dynamics\label{II}}
\subsection{Hamiltonian}
We consider two harmonic systems with coordinates and momenta ${Q}_\mu$,
${P}_\mu$, $\mu=x,y$, coupled through their angular momentum
${L}_z={Q}_x{P}_y-{Q}_y{P}_x$. The Hamiltonian is
\begin{eqnarray}
H&=&H_0-\Omega L_z\,, \label{H}\\
H_0&=&\frac{P_x^2+P_y^2}{2m} +
\frac{1}{2}(K_x {Q}_x^2+K_y {Q}_y^2)\,.\label{H0}
\end{eqnarray}
Eq.\ (\ref{H}) describes, for instance, the motion in the $x,y$ plane of a
particle of charge $e$ and mass $m$ within a harmonic trap of spring constants
$\tilde{K}_\mu$ in a uniform field $\bm{H}$ along the $z$ axis
\cite{RS.80,BR.86}, if  $\Omega=\frac{e|\bm{H}|}{2mc}$ stands for half the
cyclotron frequency and  $K_{\mu}=\tilde{K}_{\mu}+m\Omega^2$.

It also determines the intrinsic motion of a particle in a harmonic trap with
constants $K_\mu$ which rotates around the $z$ axis with frequency $\Omega$. In
this case \cite{RS.80,BR.86}, the actual Hamiltonian is $H(t)=R(t)H_0
R^\dagger(t)$, with $R(t)=e^{-i\Omega L_z t/\hbar}$ the rotation operator, but
averages of rotating observables $O(t)=R(t)OR^\dagger(t)$ evolve like those of
$O$ under the time-independent ``cranked'' Hamiltonian (\ref{H}).

Replacing ${Q}_{\mu}=q_{\mu}/\sqrt{m\Omega_0/\hbar}$,  ${P}_{\mu}=
p_{\mu}\sqrt{\hbar m\Omega_0}$, with $q_{\mu}$, $p_{\mu}$ dimensionless
coordinates and momenta ($[q_\mu,p_\nu]=i\delta_{\mu\nu}$,
$[q_\mu,q_\nu]=[p_\mu,p_\nu]=0$) and $\Omega_0$ a reference frequency,  we have
$H=\hbar\Omega_0\, h$, with
\begin{eqnarray}
h&=&h_0-\omega l_z,\;\;h_0=\frac{1}{2}(p_x^2+p_y^2+k_x q_x^2+k_y q_y^2)\label{H1a}\,,\\
l_z&=&q_x p_y-q_yp_x=-i(b^\dagger_x b_y-b^\dagger_y b_x)\,, \label{H1b}
\end{eqnarray}
where $k_{\mu}=K_{\mu}/(m\Omega_0^2)$ and $\omega=\Omega/\Omega_0$ are
dimensionless ($\Omega_0$ can be used to set $|k_x|=1$) and
$b_\mu=\frac{q_\mu+ip_\mu}{\sqrt{2}}$ are the boson annihilation operators
associated with $q_\mu$, $p_\mu$. The $l_z$ coupling (\ref{H1b}) is then seen to conserve
the associated total boson number $N=\sum_{\mu=x,y} b^\dagger_\mu b_\mu$, being
in fact the same as that describing the mixing of two modes of radiation field
passing through a beam splitter \cite{NC.00}. Notice, however, that
$[h_0,N]\neq 0$ unless $k_x=k_y=1$ (stable isotropic trap).

\subsection{Exact evolution}
The Heisenberg equations of motion $i d o/dt=-[h,o]$ for the operators $q_\mu,
p_\mu$ (with $t=\Omega_0 T$ and $T$ the actual time) become
\begin{equation}
\begin{array}{lcl}
\frac{dq_x}{dt}=p_x+\omega q_y\,,&\;&\frac{dq_y}{dt}=p_y-\omega q_x\\
\frac{dp_x}{dt}=-k_x q_x+\omega p_y\,,&\;&\frac{d p_y}{dt}=-k_y q_y-\omega p_x
\end{array}\label{S1}\,,
\end{equation}
 and can be written in matrix form as
 \begin{eqnarray}
i\frac{d}{dt}{\cal O}&=&{\cal H}{\cal O}\,,\label{eqq}\\
{\cal O}&=&\left(\begin{array}{c}q_x\\q_y\\p_x\\p_y\end{array}\right)
\,,\;\;{\cal H}=i\left(\begin{array}{cccc}
0&\omega&1&0\\-\omega&0&0&1\\-k_x&0&0&\omega\\0&-k_y&-\omega&0\end{array}\right)\,.
 \label{S2}
\end{eqnarray}
The system dynamics is then fully determined by the matrix ${\cal H}$. We may write
the general solution of (\ref{eqq})  as
\begin{equation}
{\cal O}(t)={\cal U}(t){\cal O}\,,\;\;\; {\cal U}(t)=\exp[-i{\cal H}t]\,,
 \label{Qt}\end{equation}
where ${\cal O}\equiv {\cal O}(0)$.

In spite of their simplicity,  Eqs.(\ref{S1}) can lead to  quite distinct
dynamical regimes according to the values of $\omega$ and $k_\mu$, as the
eigenvalues of ${\cal H}$, which is in general a non-hermitian matrix, can become
imaginary or complex away from stable regions \cite{RK.09}.  Moreover, ${\cal
H}$ can also become non-diagonalizable at the boundaries between distinct
regimes, exhibiting non-trivial Jordan canonical forms \cite{RK.09}.
Nonetheless, as
\begin{equation}
{\cal H}^2=\left(\begin{array}{cccc}k_x+\omega^2&0&0&-2\omega\\
0&k_y+\omega^2&2\omega&0\\
0&\omega(k_x+k_y)&k_x+\omega^2&0\\
-\omega(k_x+k_y)&0&0&k_y+\omega^2\end{array}\right),\label{H2}
\end{equation}
the eigenvalues of ${\cal H}$ are determined by $2\times 2$ blocks, and given
by  $\lambda_{\pm}$ and $-\lambda_{\pm}$, with
\begin{equation}
\lambda_{\pm}=\sqrt{\varepsilon_++\omega^2\pm\Delta}\,,
\,,\label{la}
\end{equation}
where $\varepsilon_{\pm}=\frac{k_x\pm k_y}{2}$ and
$\Delta=\sqrt{\varepsilon_-^2+4\omega^2\varepsilon_+}$.

We can then write the solution (\ref{Qt}) explicitly as
\begin{eqnarray}
\left(\begin{array}{c}q_x(t)\\ q_y(t)\\p_x(t)\\p_y(t)
\end{array}\right)
&=&\left(\begin{array}{cccc}u_{xx}&u_{xy}&v_{xx}&v_{xy}\\
u_{yx}&u_{yy}&-v_{xy}&v_{yy}\\w_{xx}&w_{xy}&u_{xx}&-u_{yx}\\
-w_{xy}&w_{yy}&-u_{xy}&u_{yy}\end{array}\right)
\left(\begin{array}{c}q_x\\ q_y\\p_x\\p_y\end{array}\right)
\,,\label{Ut}
\end{eqnarray}
where
\begin{equation}
\begin{array}{rcl}u_{\overset{xx}{_{yy}}}&=&\frac{(\Delta\pm\varepsilon_-)
\cos\lambda_{+}t+(\Delta\mp \varepsilon_-)\cos\lambda_- t}{2\Delta}\, ,\\
u_{\overset{xy}{_{yx}}}&=&\pm\omega\frac{(\Delta\mp\varepsilon_-+2\varepsilon_+)
\frac{\sin\lambda_+ t}{\lambda_+}+
(\Delta\pm\varepsilon_--2\varepsilon_+)\frac{\sin\lambda_- t}{\lambda_-}}{2\Delta} \, ,\\
v_{\overset{xx}{_{yy}}}&=
&\frac{(\Delta\pm\varepsilon_-+2\omega^2)\frac{\sin\lambda_+ t}{\lambda_+}+
(\Delta\mp\varepsilon_--2\omega^2)\frac{\sin\lambda_- t}{\lambda_-}}{2\Delta}\, ,\\
v_{xy}&=&\frac{\omega(-\cos\lambda_+ t+\cos\lambda_- t)}{\Delta}\,,\;\;
w_{xy}=-\varepsilon_+v_{xy}\, ,\\
w_{\overset{xx}{_{yy}}}&=& \frac{-(\Delta\pm\varepsilon_-)
(\Delta\pm\varepsilon_-+2\varepsilon_+)\frac{\sin\lambda_+t}{\lambda_+} +
(\Delta\mp\varepsilon_-)(\Delta\mp\varepsilon_--2\varepsilon_+)\frac{\sin\lambda_-
t}{\lambda_-}} {4\Delta}\, .
\end{array} \label{sub}
\end{equation}
The matrix  ${\cal U}(t)$ is real  for any real values
of $\omega$, $k_\mu$ and $t$, including unstable regimes where $\Delta$ and/or
$\lambda_\pm$ can be imaginary or complex \cite{RK.09}. It represents always a
linear canonical transformation of the $q_\mu$, $p_\mu$,  satisfying
\begin{equation}
{\cal U}(t){\cal M}{\cal U}^{t}(t)={\cal M},\;\;
{\cal M}=i\left(\begin{array}{cc}0&I\\-I&0\end{array}\right)\,,
\label{M}
\end{equation}
($I$ denotes the $2\times 2$ identity matrix) which ensures the preservation of
commutation relations ($[{\cal O}_i,{\cal O}_j]={\cal M}_{ij}$). It corresponds to a
proper Bogoliubov transformation of the associated boson operators.

For $\omega=0$, we recover from Eqs.\ (\ref{Ut})--(\ref{sub}) the decoupled
harmonic evolution $q_\mu(t)=q_\mu\cos\omega_\mu t+\omega_\mu^{-1}p_\mu
\sin\omega_\mu t$, $p_\mu(t)=p_\mu\cos\omega_\mu
t-q_\mu\omega_\mu\sin\omega_\mu t$, where $\omega_\mu=\sqrt{k_\mu}$ for
$\mu=x,y$.  Off-diagonal terms $u_{xy}$, $u_{yx}$, $v_{xy}$, $w_{xy}$ in
(\ref{sub}) are $O(\omega)$ for small $\omega$.

On the other hand, in the isotropic case  $k_x=k_y=k$ (where $\Delta=2\omega
\sqrt{k}$ and $|\lambda_{\pm}|=|\sqrt{k}\pm\omega|$), $[l_z,h]=0$ and the evolution
provided by Eqs.\ (\ref{Ut})--(\ref{sub}) is just the rotation of  identical
single mode evolutions:
\begin{eqnarray}{\cal U}(t)&=&\exp[i\omega{\cal L}_z t]\exp[-i{\cal H}_0t]\,,\label{rot}\\
\exp[i\omega{\cal L}_z t]&=&\left(\begin{array}{cc}R^\dagger(t)&0\\0&R^\dagger(t)
\end{array}\right),\;R^\dagger(t)=\left(\begin{array}{cc}\cos \omega t&\sin\omega t\\
-\sin\omega t&\cos\omega t\end{array}\right)
.\nonumber\end{eqnarray}
In particular, the  Landau case (free particle in a magnetic field)
corresponds to $k_x=k_y=\omega^2$, where $\lambda_+=2\omega$ and $\lambda_-=0$.

\subsection{Dynamical regimes}
\begin{figure}[t] \centerline{
\hspace{1.cm}\includegraphics[trim=0cm 0cm 0cm 0cm, width = 8cm]{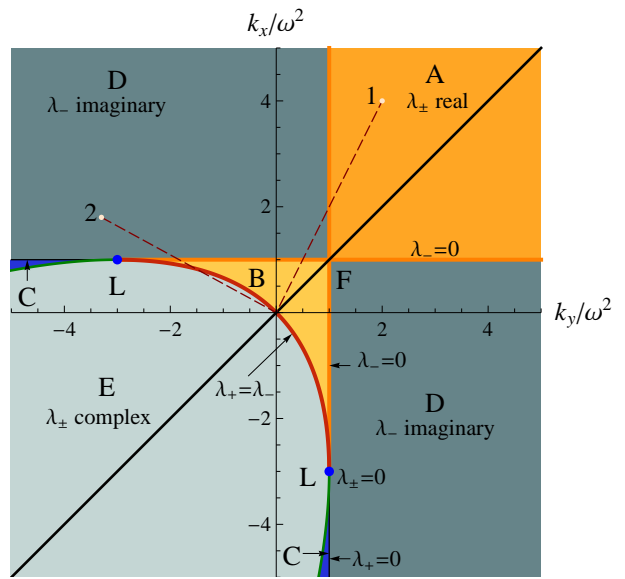}}
\vspace*{0 cm} \caption{Dynamical phase diagram of the system described by
Hamiltonian (\ref{H}). The evolution of the operators $q_\mu$, $p_\mu$ is
quasiperiodic in the dynamically stable sectors  {\bf A}, {\bf B}, where the
eigenfrequencies $\lambda_{\pm}$ are both real, but unbounded in the remaining
sectors, with $\lambda_{-}$ imaginary in  ${\bf D}$,  both $\lambda_{\pm}$
imaginary in ${\bf C}$, and $\lambda_{\pm}$ complex conjugates in ${\bf E}$. At
the borders between these regions (except from the Landau point $F$) the matrix
${\cal H}$ is non-diagonalizable and the evolution is also unbounded, with
$\lambda_-=0$ at the borders between {\bf D} and {\bf A} or {\bf B},
$\lambda_+=0$ at the border between {\bf D} and {\bf C},  $\lambda_+=\lambda_-$
at the curve separating {\bf E} from {\bf B} and {\bf C} and
$\lambda_{\pm}=0$ at the critical points $L$. Dashed lines indicate the path
described as $\omega$ is increased at fixed $k_\mu$, showing that the system
dynamics will become unbounded (bounded) in a certain frequency window when
starting at $1$ ($2$). The black line indicates the isotropic case $k_x=k_y$,
where entanglement will be periodic (see sec.\ \ref{III}).} \label{f1}
\end{figure}

The distinct dynamical regimes exhibited by this system for $\omega\neq 0$ are
summarized in Fig.\ \ref{f1}. Let us first consider the standard stable case
$k_x>0$, $k_y>0$ (first quadrant). The eigenvalues $\lambda_{\pm}$ are here
both real and non-zero in sectors {\bf A}  and {\bf B}, defined by
\begin{eqnarray}\omega^2&<&{\rm Min}[k_x,k_y]\;\;\;\;\;
({\rm sector}\;{\rm\bf A})\,,\label{omega1}\\
\omega^2&>&{\rm Max}[k_x,k_y]\;\;\;\;\; ({\rm sector}\;
 {\rm\bf B})\,,\label{omega2}\end{eqnarray}
when  $k_x>0$, $k_y>0$. {\bf A} is the full stable sector where $h$ is positive
definite, whereas {\bf B} is that where the system, though unstable, remains
dynamically stable \cite{RK.09} (see also Appendix). If $\omega^2$ lies between
these values (sector {\bf D}), $\lambda_-$ becomes {\it imaginary} (with
$\lambda_+$ remaining real), leading to a frequency window where the system
becomes dynamically unstable (unbounded motion), with  $\sin(\lambda_-
t)/\lambda_-=\sinh(|\lambda_-|t)/|\lambda_-|$ in Eqs.\ (\ref{sub}).

At the border between  {\bf D} and {\bf A} or {\bf B} ($\omega^2=k_x$ or
$\omega^2=k_y$), $\lambda_-=0$ (with $\lambda_+>0$) and ${\cal H}$ becomes
non-diagonalizable if $k_y\neq k_x$, although ${\cal H}^2$ remains
diagonalizable. The system becomes here equivalent to a stable oscillator plus
a free particle \cite{RK.09} (see Appendix), and we should just replace
$\sin(\lambda_- t)/\lambda_-$ by its limit $t$  in Eqs.\ (\ref{sub}), which
leads again to an unbounded motion.

 \vspace*{-0.cm}

Considering now the possibility of unstable potentials ($k_x<0$ and/or $k_y<0$,
remaining quadrants), the dynamically stable sector {\bf B} extends into this
region provided $k_x>0>k_y>-3k_x$ (or viceversa) and
\begin{equation}
{\rm Max}[k_x,k_y]<\omega^2<-\varepsilon^2_-/(4\varepsilon_+)\,,\label{omega3}
\end{equation}
where the upper bound applies only when $\varepsilon_+<0$ (i.e.,
$-3k_x<k_y<-k_x$ or viceversa). Eq. (\ref{omega3}) defines a frequency window
where the unstable system becomes dynamically stable ($\lambda_{\pm}$ real).
Beyond this sector, either $\lambda_-$ becomes imaginary (sectors ${\bf D}$) or
both $\lambda_{\pm}$ become imaginary (sectors {\bf C}) or complex conjugates
(sector {\bf E}, where $\Delta$ is imaginary),  and the dynamics becomes again
unbounded. This is also the case at the borders between {\bf D} and {\bf B}
($\lambda_+>0$, $\lambda_-=0$) and also {\bf D} and {\bf C} ($\lambda_+=0$,
$\lambda_-$ imaginary) where ${\cal H}$ is non-diagonalizable (see Appendix for
more details).

The critical curve $\Delta=0$, i.e.,
\begin{equation}\omega^2=-\varepsilon^2_-/(4\varepsilon_+)
 \,,\label{wc}\end{equation}
where $\varepsilon_+<0$, separates sectors {\bf B} and {\bf C} from {\bf E} and
deserves special attention. At this curve,
$\lambda_{\pm}=\lambda=\sqrt{\varepsilon_++\omega^2}$ and both ${\cal H}$ {\it
and} ${\cal H}^2$ become {\it non-diagonalizable}, with $\lambda$ real at the
border between {\bf B} and {\bf E} and imaginary at that between {\bf C} and
{\bf E}. The evaluation of ${\cal U}(t)$ in Eq.\ (\ref{Qt}) can in this case be
obtained through the pertinent Jordan decomposition of ${\cal H}$ (two $2\times
2$ blocks \cite{RK.09}), but  the final result coincides with the
$\Delta\rightarrow 0$ limit of Eqs.\ (\ref{sub}). This leads to the elements
\begin{equation}\label{sub2}\begin{array}{rcl}
u_{\overset{xx}{_{yy}}}&=&\cos\lambda t\mp t\varepsilon_-\frac{\sin\lambda t}{2\lambda}\,,\\
u_{\overset{xy}{_{yx}}}&=&\pm\omega\frac{t\lambda(\varepsilon_+\mp\varepsilon_-/2)\cos\lambda t+
(\omega^2\pm\varepsilon_-/2)\sin\lambda t}{\lambda^3}\,,\\
v_{\overset{xx}{_{yy}}}&=&\frac{t\lambda(\omega^2\pm\varepsilon_-/2)\cos\lambda t+
(\varepsilon_+\mp\varepsilon_-/2)\sin\lambda t}{\lambda^3}\,,\\
v_{xy}&=&\omega t\frac{\sin\lambda t}{\lambda}\,,\;\;w_{xy}=-\varepsilon_+ v_{xy}\,,\\
w_{\overset{xx}{_{yy}}}&=&\frac{\varepsilon_+ t\lambda(\omega^2\mp\varepsilon_-/2)\cos\lambda t-
(\varepsilon_++2\omega^2)(\varepsilon_+\pm\varepsilon_-/2)\sin\lambda t}{\lambda^3}\,,
 \end{array}
\end{equation}
which contain terms proportional to $t$. The evolution is, therefore, always
unbounded along this curve.

Finally, if both $\Delta$ {\it and} $\lambda=\sqrt{\varepsilon_++\omega^2}$
vanish, which occurs when  $\varepsilon_+=-\omega^2=-|\varepsilon_-|/2$, i.e.,
\begin{equation}\omega^2=k_x=-k_y/3\,,\label{wcc}\end{equation}
(or $\omega^2=k_y=-k_x/3$), the system exhibits a remarkable {\it critical
point} (points $L$ in Fig.\ \ref{f1}), where $\lambda_{\pm}=0$ and sectors {\bf
B}, {\bf C}, {\bf D} and {\bf E} meet. Here both ${\cal H}$ and ${\cal H}^2$
are non-diagonalizable, with ${\cal H}$ represented by a single $4\times 4$
Jordan Block (inseparable pair \cite{RK.09}). By using this form or taking the
$\lambda\rightarrow 0$ limit in Eqs.\ (\ref{sub2}), we obtain in this case a
purely {\it polynomial} (and hence also unbounded) evolution, involving terms
up to the {\it third} power of $t$: The elements of ${\cal U}(t)$ become
\begin{equation}\label{sub3}\begin{array}{rclrcl}
u_{\overset{xx}{_{yy}}}&=&1\mp \omega^2t^2&&&\\
u_{xy}&=&\omega t(1+\frac{2}{3}\omega^2 t^2),&
u_{yx}&=&-\omega t\,,\\
v_{xx}&=&t(1-\frac{2}{3}\omega^2t^2),&
v_{yy}&=&t\,,\\v_{xy}&=&\omega t^2\,,&w_{xy}&=&\omega^3t^2\,,\\
w_{xx}&=&-\omega^2 t,&w_{yy}&=&\omega^2 t(3+\frac{2}{3}\omega^2t^2)\,. \end{array}
\end{equation}
Nonetheless, we remark that Eq.\ (\ref{M}) remains satisfied (in both cases
(\ref{sub2}) and (\ref{sub3})).
\section{Dynamics of entanglement in gaussian states\label{III}}
\subsection{Exact evaluation}
Let us now consider the evolution of the entanglement between the $x$ and $y$
modes, starting from an initially separable pure gaussian state. Since the
evolution is equivalent to the linear canonical transformation (\ref{Qt}), the
state will remain gaussian $\forall$ $t$, which entails that entanglement will
be completely determined by the pertinent covariance matrix \cite{AEPW.02,ASI.04}.

We may then assume that at $t=0$, $\langle q_\mu\rangle=\langle p_\mu\rangle=0$
for $\mu=x,y$ ($\langle{\cal O}\rangle=0$), such that these mean values will
vanish $\forall$ $t$ ($\langle {\cal O}\rangle_t=\langle {\cal O}(t)\rangle=0$,
as implied by Eq.\ (\ref{Ut})). We may then define the  covariance matrix as
\begin{eqnarray}
{\cal C}&=&\langle {\cal O}{\cal O}^t\rangle-\frac{1}{2}{\cal M}
\nonumber\\
&=&
\left(\begin{array}{cccc}\langle q_x^2\rangle &\langle q_x q_y\rangle&
\frac{\langle q_x p_x+p_xq_x\rangle}{2}&\langle q_x p_y\rangle\\
\langle q_x q_y\rangle&\langle q_y^2\rangle&\langle q_y p_x
\rangle&\frac{\langle q_y p_y+p_y q_y\rangle}{2}\\\frac{\langle q_x p_x+p_xq_x
\rangle}{2}&\langle q_y p_x\rangle
&\langle p_x^2\rangle &\langle p_x p_y\rangle\\
\langle q_x p_y\rangle&\frac{\langle q_y p_y+p_y q_y\rangle}{2}&
\langle p_x p_y\rangle&\langle p_y^2\rangle\end{array}\right)\label{C0}\,,\end{eqnarray}
which, according to Eqs.\ (\ref{Qt}) and (\ref{M}), will evolve as
\begin{equation}{\cal C}(t)={\cal U}(t){\cal C}(0){\cal U}^t(t)\label{Ct}\,.\end{equation}

The entanglement between the two modes will now be determined by the symplectic
eigenvalue $\tilde{f}(t)=f(t)+1/2$ of the single mode covariance matrix ${\cal
C}_\mu(t)=\langle {\cal O}_\mu {\cal O}_\mu^t\rangle_t-\frac{1}{2}{\cal M}$,
submatrix of (\ref{Ct}), where ${\cal O}_\mu=(q_\mu,p_\mu)^t$. Here $f(t)$ is a
non-negative quantity representing the average boson occupation $\langle
a^\dagger_\mu(t) a_\mu(t)\rangle$ of the mode ($a_\mu(t)$ is the local boson
operator satisfying $\langle a^2_\mu(t)\rangle=0$),  which is the same for both
modes ($f_x(t)=f_y(t)$) when the global state is gaussian and pure. It is given
by
\begin{equation}
f(t)=\sqrt{\langle q_\mu^2\rangle_t\langle p_\mu^2\rangle_t-\langle q_\mu
 p_\mu+p_\mu q_\mu\rangle_t^2/4}-\frac{1}{2} \label{fmu}\,. \end{equation}
Eq.\ (\ref{fmu}) is just the deviation from minimum uncertainty of the mode,
and can be directly determined from the elements of (\ref{Ct}).

The von Neumann entanglement entropy between the two modes becomes 
\begin{eqnarray}S(t)&=&-{\rm Tr}\,\rho_\mu(t)\ln \rho_\mu(t)\nonumber\\
&=&-f(t)\ln f(t)+[1+f(t)]\ln[1+f(t)]\,,\label{Smu}
\end{eqnarray}
where $\rho_\mu(t)$ denotes  the reduced state of the
mode. Eq.\ (\ref{Smu}) is an increasing concave function of $f(t)$.
For future use, we note that for large and small $f(t)$,
\begin{eqnarray}S(t) &\approx&\ln f(t)+1+
O(f^{-1})\,,\label{finf}\\ S(t)&\approx& f(t)[-\ln f(t)+1]+O(f^2)\,.\label{f0}
\end{eqnarray}
Other entanglement entropies, like the Renyi entropies $S_\alpha(t)=\frac{\ln
{\rm Tr}\,\rho_\mu^\alpha(t)}{1-\alpha}$, $\alpha>0$, and the linear entropy
$S_2(t)=1-{\rm Tr}\rho_\mu^2(t)$ (of experimental interest as ${\rm Tr}\rho^2$
and in general ${\rm Tr}\rho^n$ can be measured without performing a full state
tomography \cite{DPSZ.12,EA.02}), are obviously also determined by $f(t)$,
since ${\rm Tr\,}\rho_\mu^\alpha=[(1+f_\mu)^\alpha-f_\mu^\alpha]^{-1}$
($\alpha>0$).

The initial covariance matrix ${\cal C}(0)$ will be here assumed of the form
\begin{equation}{\cal C}(0)=\frac{1}{2}\left(\begin{array}{cccc}\alpha_x^{-1} &0&0&0\\
0&\alpha_y^{-1}&0&0\\0&0&\alpha_x&0\\0&0&0&\alpha_y\end{array}\right)\,,
 \label{C00}\end{equation}
where $\alpha_\mu=2\langle p_\mu^2\rangle$, such that
$\alpha_\mu=\sqrt{k_{\mu}}$ if the system is initially in the separable ground
state of $h_0$, as in the typical quantum quench scenario \cite{SLRD.13}. For
fixed isotropic initial conditions we will just take $\alpha_x=\alpha_y=1$.

For these initial conditions, we first notice that for small $t$,
Eqs.\ (\ref{sub}) and (\ref{fmu}) yield
\begin{equation}f(t)\approx
\frac{(\alpha_x-\alpha_y)^2}{4\alpha_x\alpha_y}(\omega t)^2+O(t^4)\,,
 \label{fi}\end{equation}
which indicates a quadratic initial increase of $f(t)$ with time for any
anisotropic initial covariance. Eq.\ (\ref{fi}) is independent of the
oscillator parameters $k_\mu$ and proportional to $\omega^2$. However, for {\it
isotropic} initial conditions $\alpha_x=\alpha_y$, quadratic terms vanish and
we obtain instead a quartic initial increase, driven by the oscillator
anisotropy $\varepsilon_-$:
\begin{equation}
f(t)\approx
 \frac{\varepsilon_-^2\omega^2}{4\alpha_x^2}t^4+O(t^6)\,.\end{equation}
 Eq.\ (\ref{f0}) implies a
similar initial behavior (except for a factor $\ln t$) of the entanglement
entropy.

Next, in the isotropic case $k_x=k_y=k$, the exact expression for
$f(t)$ becomes quite simple, since the rotation is decoupled from the internal
motion of the modes (Eq.\ (\ref{rot})), and entanglement arises {\it solely
from rotation and initial anisotropy}. We obtain
\begin{equation}
f(t)=\frac{1}{2}\sqrt{1+\frac{(\alpha_x-\alpha_y)^2}{4\alpha_x\alpha_y}
 \sin^2(2\omega t)}-\frac{1}{2}\,.\label{isot}\end{equation}
Entanglement will then simply oscillate with frequency $4\omega$ if
$\alpha_x\neq\alpha_y$, being {\it independent} of the trap parameter $k$,
since the latter affects just a local transformation decoupled from the
rotation. Eq.\ (\ref{isot}) holds in fact even if $k$ becomes negative
(unstable potential) or vanishes.

In the general case, the previous decoupling no longer holds and the explicit
expression for $f(t)$ becomes quite long. The main point we want to show is
that the different dynamical regimes lead to distinct behaviors of $f(t)$, and
hence of the generated entanglement entropy $S(t)$, which are summarized in
Table \ref{t1}. We now describe them in detail.

\begin{table}[t]
\centerline{\hspace*{0.cm}\includegraphics[trim=0cm 0cm 0cm 0cm, width =
 8cm]{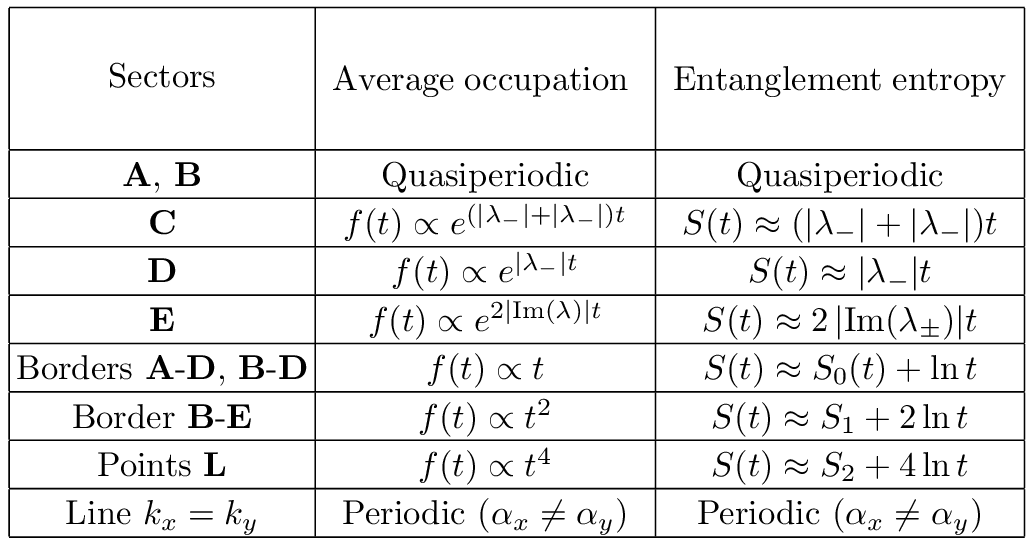}}
\caption{The asymptotic evolution of the average occupation (\ref{fmu}) and the
entanglement entropy (\ref{Smu}) in the different dynamical sectors indicated
in Fig.\ \ref{f1}. Entanglement is bounded in the stable sectors {\bf A}, {\bf
B}, but increases linearly (with $t$) in the unstable sectors {\bf C}, {\bf D},
{\bf E}, and logarithmically  at the border between stable and unstable
sectors, provided $k_x\neq k_y$.  In the isotropic case  $k_x=k_y=k$
 it remains periodic for any value of $k$ and anisotropic initial conditions.} \label{t1}
\end{table}
\vspace*{-0.cm}

\subsection{Evolution in stable sectors}
\begin{figure}[t]
\centerline{\hspace*{0.cm}\includegraphics[trim=0cm 0cm 0cm 0cm, width =
 9cm]{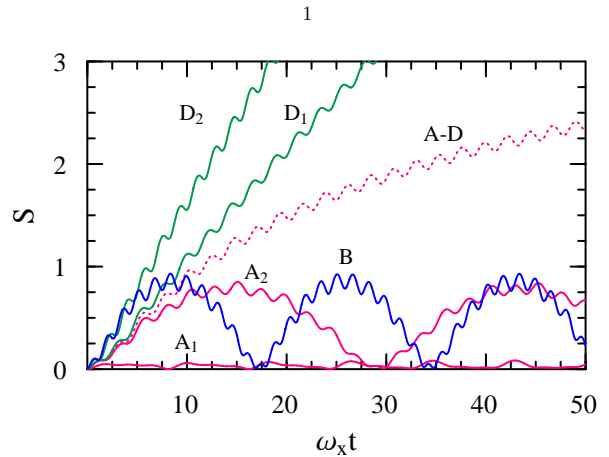}}
\caption{The evolution of the entanglement entropy (\ref{Smu}) between the two
modes for $k_y=0.3 k_x>0$ and frequencies $\omega/\omega_y=0.5$ ({\bf A}$_1$),
$0.95$ ({\bf A}$_2$),  $1$ ({\bf A-D}), $1.05$ ({\bf D}$_1$), $1.7$ ({\bf
D}$_2$) and $1.95$ ({\bf B}), where $\omega_\mu=\sqrt{k_\mu}$ and the label
indicates the corresponding sector in Fig.\ \ref{f1}. $S(t)$ is quasiperiodic
in curves {\bf A}$_1$, {\bf A}$_2$ and {\bf B}, but increases logarithmically
(on average) in {\bf A}-{\bf D}, and linearly in {\bf D}$_1$, {\bf D}$_2$. The
initial state is the separable ground state of $H_0$ (uncoupled oscillators).}
\label{f2}
\end{figure}

In the dynamically stable sectors {\bf A} and {\bf B}  of  Fig.\ \ref{f1},
both $\lambda_{\pm}$ are real and non-zero, implying that the evolution
of $f(t)$ and $S(t)$ will be quasiperiodic, as seen in Fig.\ \ref{f2} (curves
{\bf A}$_1$, {\bf A}$_2$ and {\bf B}). The initial state was chosen as the
ground state of $h_0$ ($\alpha_\mu=\sqrt{k_\mu}$ in (\ref{C00})). Starting from
point $1$ in sector {\bf A} (Fig.\ \ref{f1}), the generated entanglement $S(t)$
remains small when $\omega$ is well below the first critical value
$\omega_y=\sqrt{k_y}$ (curve {\bf A}$_1$). As $\omega$ increases,  $S(t)$ will
exhibit increasingly higher maxima, showing a typical resonant behavior for
$\omega$ close to  $\omega_y$ (border with sector {\bf D}), where $\lambda_-$
vanishes. Near this border, $S(t)$ will essentially exhibit large amplitude low
frequency oscillations determined by $\lambda_-$, with maxima at $t\approx t_m=
\frac{m\pi}{2\lambda_-}$ ($m$ odd), plus low amplitude high frequency
oscillations determined by $\lambda_+$, as seen in curve {\bf A}$_2$.

\vspace*{-0.cm}

As $\omega$ increases, the system enters dynamically unstable sectors for
$\omega_y\leq \omega\leq\omega_x=\sqrt{k_x}$, and the evolution becomes
unbounded (curves {\bf A-D}, {\bf D}$_1$ and {\bf D}$_2$, described in next
subsection). For  $\omega>\omega_x$, the system reenters the dynamically stable
regime and exhibits again the previous behaviors, with an oscillatory resonant
type evolution for $\omega$ above but close to $\omega_x$ (curve {\bf B} in
Fig.\ \ref{f2}).

\begin{figure}[t]
\centerline{\hspace*{0.cm}\includegraphics[trim=0cm 0cm 0cm 0cm, width
 =7cm]{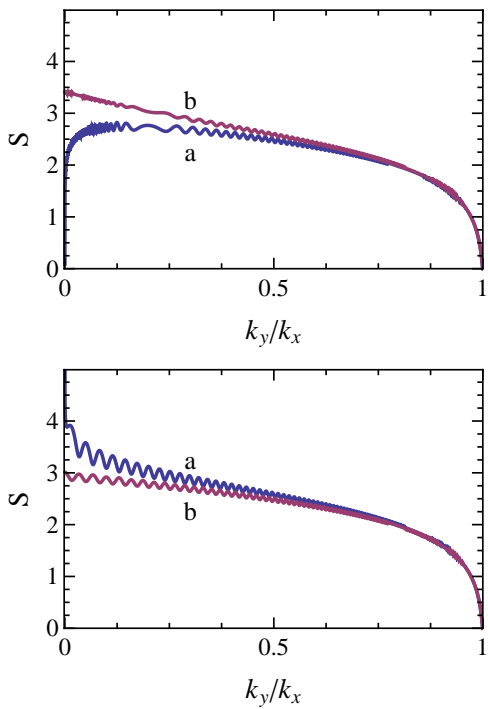}}
\caption{The maximum entanglement $S(t_m)$ reached in stable sectors close to
instability, as a function of the anisotropy ratio $k_y/k_x$ (see Eq.\
(\ref{flineal2})). Top: Vicinity of border {\bf A}-{\bf D}
($\omega=0.999\omega_y$). Bottom: Vicinity of border {\bf B}-{\bf D}
($\omega=1.001\omega_x$). The initial state is the separable ground state of
$H_0$ ($\alpha_\mu=\omega_\mu$) in curves {\bf a}  and a separable isotropic
state ($\alpha_\mu=1$) in curves {\bf b}.} \label{f3}
\end{figure}
\vspace*{-0.cm}

Close to instability but still within the stable regime, the maximum
entanglement reached is of order $\ln|\omega-\omega_\mu|$: For $\omega$ close
to $\omega_\mu$ ($\mu=x,y$) on the stable side, and for the initial conditions
(\ref{C00}), $f(t)$ will be maximum at $t\approx t_m $, with
\begin{equation}
f(t_m)\approx{\textstyle\frac{\omega |\varepsilon_-|}
{\lambda_+^2\lambda_-}\sqrt{\frac{
(\frac{\alpha_x\alpha_y+\omega^2}{\lambda_+})^2\sin^2\frac{m\pi\lambda_+}
{2\lambda_-}+\alpha_\mu^2\cos^2\frac{m\pi\lambda_+}{2\lambda_-}}
{\alpha_x\alpha_y}}}\,,\label{flineal2}
\end{equation}
where $\lambda_+\approx \sqrt{2(\varepsilon_++\omega_\mu^2)}$ and
\begin{equation}\lambda_-\approx\sqrt{\frac{2\omega_\mu|\varepsilon_-|
|\omega_\mu-\omega|}{\varepsilon_++\omega_\mu^2}}
 \,,\label{lm}\end{equation}
implying $f(t_m)=O(|\omega_\mu-\omega|^{-1/2})$ and hence
$S(t_m)=O(-\frac{1}{2}\ln|\omega_\mu-\omega|)$.

Expression (\ref{flineal2}) (and hence $S(t_m)$) will tend to decrease for
decreasing anisotropy, i.e., increasing ratio $k_y/k_x\leq 1$, as seen in Fig.\
\ref{f3} for $m=1$, vanishing in the isotropic limit $k_y/ k_x\rightarrow 1$
(where $f(t_m)=O(|k_x-k_y|)^{1/2})$. On the other hand, the behavior for
$k_y/k_x\rightarrow 0$ will depend on the initial condition: If it is the
ground state of $H_0$ ($\alpha_\mu=\omega_\mu$, curves {\bf a}), $f(t_m)$ will
vanish at the first border $\omega\approx\omega_y$ (top panel), where
$f(t_m)=O(\sqrt{\omega_y})$, but diverge at the second border $\omega=\omega_x$
(bottom panel), where $f(t_m)=O(1/\sqrt{\omega_y})$, as obtained from  Eq.\
(\ref{flineal2}). If the initial state is fixed, however, $f(t_m)$ will
approach a finite value for $k_y/k_x\rightarrow 0$, and exhibit a monotonous
decrease on average with increasing ratio $k_y/k_x$ in both borders (curves
{\bf b} in Fig.\ \ref{f3}), as also implied by (\ref{flineal2}). We also
mention that the high frequency oscillations in $f(t_m)$ and $S(t_m)$ observed
in Fig.\ \ref{f3} stem from the $\lambda_+/\lambda_-$ ratio in the arguments of
the trigonometric functions in Eq.\ (\ref{flineal2}). For $\omega$ close to
$\omega_\mu$, this ratio is minimum around $k_y/k_x\approx 1/5$, which leads to
the observed decrease in the oscillation frequency of $S(t_m)$ in the vicinity
of this ratio (top panel).

\subsection{Evolution in unstable sectors} Let us now examine in detail the
evolution of $S(t)$ in the dynamically unstable regimes. At the critical
frequencies $\omega=\omega_\mu$, $\mu=y,x$ (borders {\bf A}-{\bf D} and {\bf
B}-{\bf D}), $\lambda_-$ vanishes and Eqs.\ (\ref{sub}) and (\ref{fmu}) lead,
for large $t$ and the initial conditions (\ref{C00}), to the critical evolution
\begin{equation}
f(t)\approx t\,{\textstyle\frac{\omega |\varepsilon_-|}
{\lambda_+^2}\sqrt{\frac{
(\frac{\alpha_x\alpha_y+\omega^2}{\lambda_+})^2\sin^2 \lambda_+ t
+\alpha_\mu^2\cos^2\lambda_+ t}
{\alpha_x\alpha_y}}}\,,\label{flineal}
\end{equation}
where $\lambda_+=\sqrt{2(\varepsilon_++\omega_\mu^2)}>0$. This entails a {\it linear}
increase, on average,  of $f(t)$ in this limit, and hence, a {\it
logarithmic} growth  of $S(t)$, according to Eq.\
(\ref{finf}):
\begin{equation}
 S(t)\approx S_0(t)+\ln t\,, \label{as1}\end{equation}
where $S_0(t)=1+\ln [f(t)/t]$ is a bounded function oscillating with frequency
$\lambda_+$. This behavior (curve {\bf A}-{\bf D} in Fig.\ \ref{f2}) is the
$\omega\rightarrow\omega_\mu$ limit of the previous resonant regime.

\begin{figure}[t]
\centerline{\hspace*{0.cm}\includegraphics[trim=0cm 0cm 0cm 0cm, width =
 7cm]{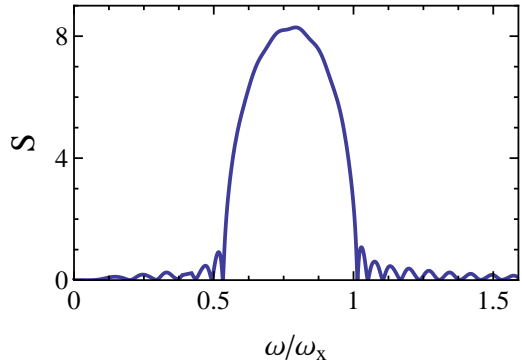}}
\caption{The entanglement entropy  between the two modes  attained at fixed
time $\omega_x t=40$, as a function of the (constant) frequency $\omega$, for
the oscillator parameters and initial state of Fig.\ \ref{f2}. Entanglement is
bounded for $\omega<\omega_y$ (sector {\bf A}) and $\omega>\omega_x$ (sector
{\bf B}), but is proportional to $t$ in the instability window
$\omega_y<\omega<\omega_x$ (sector {\bf D}).} \label{f4}
\end{figure}

On the other hand, in the unstable sector {\bf D} ($\omega_y<\omega<\omega_x$),
$\lambda_-$ becomes imaginary. This leads to an {\it exponential} term in
$f(t)$ ($\frac{\sin\lambda_- t}{\lambda_-}\rightarrow \frac{\sinh|\lambda_-|
t}{|\lambda_-|}$), which will dominate the large $t$ evolution: In this sector
Eqs.\ (\ref{sub}), (\ref{fmu}) and (\ref{f0}) imply, for large $t$,
\begin{equation}f(t)\propto e^{|\lambda_-|t}\,,\;\;\;S(t)\approx |\lambda_-|t\,,
\end{equation}
and hence, a {\it linear} growth (on average) of the entanglement entropy with
time (curves {\bf D}$_1$, {\bf D}$_2$ in Fig.\ \ref{f2}). Therefore, in the
unstable window $\omega_y\leq\omega\leq\omega_x$, there is an unbounded growth
with time of the entanglement entropy, which will originate a pronounced
maximum in the generated entanglement at a given fixed time and anisotropy as a
function of $\omega$, as appreciated in Fig.\ \ref{f4}.

 \vspace*{-0.cm}

We now  examine the behavior at the other sectors of Fig.\ \ref{f1}. In the
unstable sectors {\bf C} and {\bf E}, where one or both of the constants
$k_\mu$ are negative, $\lambda_{\pm}$ are imaginary or complex (Fig.\
\ref{t1}). This implies an exponential increase of $f(t)$, as indicated in
table \ref{t1}, entailing again a {\it linear} asymptotic growth of the
entanglement entropy with time: $S(t)\approx(|\lambda_+|+|\lambda_-|)t$ in {\bf
C} and $S(t)\approx 2|{\rm Im}(\lambda_{\pm})| t$ in {\bf E}, neglecting
constant or bounded terms.

On the other hand, at the border between sectors {\bf B} and {\bf E}, which
corresponds to the critical curve $\Delta=0$ between both points $L$ in Fig.\
\ref{f1}, we obtain, for large $t$ and $k_x\neq k_y$ (with the initial
conditions (\ref{C00})), the asymptotic behavior
\begin{equation}
f(t)\approx  |\varepsilon_-|
\frac{4\omega^2\alpha_x\alpha_y+\varepsilon_-^2}
{16\omega\lambda\sqrt{\alpha_x\alpha_y}}
\,t^2\,,\label{f2t}\end{equation} where
$\lambda=\sqrt{\varepsilon_++\omega^2}>0$. This leads to
\begin{equation} S(t)\approx S_1+2\ln t\,,\label{as2}\end{equation}
with $S_1\approx1+\ln[|\varepsilon_-|\frac{4\omega^2\alpha_x\alpha_y
+\varepsilon_-^2} {16\omega\lambda\sqrt{\alpha_x\alpha_y}}]$. Hence, the
unbounded growth  of $f(t)$ and $S(t)$ is here more rapid  than that at the
previous borders {\bf A}-{\bf D} and {\bf B}-{\bf D} ($\omega=\omega_y$ or
$\omega_x$) (quadratic instead of linear increase of $f(t)$). At the  border
{\bf E}-{\bf C} the asymptotic behavior of $f(t)$ is still exponential (i.e.,
linear growth of $S(t)$).

 \begin{figure}[t]
\centerline{ \hspace*{0.cm}\includegraphics[trim=0cm 0cm 0cm 0cm, width =
 7cm]{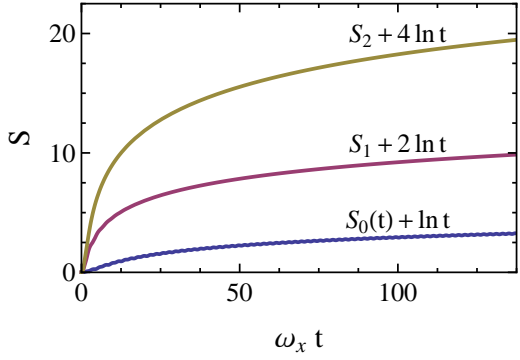}}
\caption{Critical evolution of the entanglement entropy at the border between
sectors with distinct dynamics, for isotropic initial conditions
($\alpha_\mu=1$). The lower, middle and upper curve correspond respectively to
the border {\bf A}-{\bf D} (at $k_y=0.5 k_x$, with $\omega=\sqrt{k_y}$), {\bf
B}-{\bf E} (at $k_y=-1.5 k_x$, with $\omega$ given by (\ref{wc})) and the
critical points {\bf L} (Eq.\ (\ref{wcc})). The asymptotic behavior for large
$t$ (Eqs.\ (\ref{as1}), (\ref{as2}), (\ref{as3})) is indicated.} \label{f5}
\end{figure}
\vspace*{-0.cm}

Finally, a further remarkable critical behavior arises at the
special critical points $L$, obtained for condition (\ref{wcc}), where all
sectors {\bf B, C, D} and {\bf E} meet. We obtain here a purely {\it
polynomial} evolution of $(f(t)+1/2)^2$, as implied by Eqs.\ (\ref{sub3}). For
large $t$, this leads to a {\it quartic} increase of $f(t)$:
\begin{equation}
f(t)\approx {\textstyle
\frac{\alpha_x\alpha_y+\omega^2}{6\sqrt{\alpha_x\alpha_y}}}\omega^3\, t^4\,,
\label{f4t}\end{equation}
implying the following logarithmic increase of $S(t)$:
\begin{equation} S(t)\approx S_2+4\ln t\,,\label{as3}\end{equation}
where $S_2\approx 1+\ln[\frac{\alpha_x\alpha_y+\omega^2}
{6\sqrt{\alpha_x\alpha_y}} \omega^3]$. Hence, the
increase is here still more rapid than at both previous borders. These critical
behaviors are all depicted in Fig.\ \ref{f5}.

\subsection{Entanglement control}
\begin{figure}[t] \centerline{\hspace*{0.cm}\includegraphics[trim=0cm 0cm 0cm 0cm,
 width = 7cm]{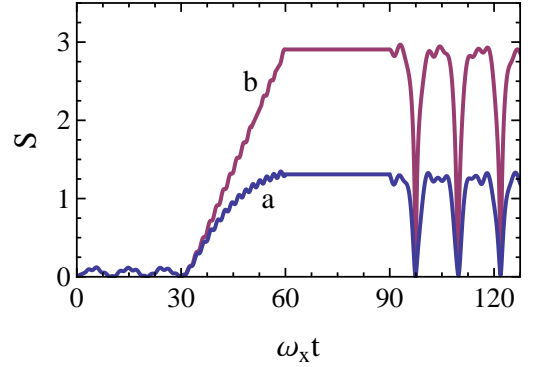}}
\caption{Evolution of the entanglement entropy for a stepwise varying frequency
$\omega$, starting from the separable ground state of $H_0$ (with $k_y=0.5
k_x>0$). In curve {\bf a} we have set $\omega/\omega_x=0.5$, $0.7$, $0$ and
$0.21$ for successive time intervals of length $\omega_x \Delta t=30$, such
that the system is close to the first instability at the second interval ($0.7
\omega_x\approx 0.99 \omega_y$, with $\omega_\mu=\sqrt{k_\mu}$), while in curve
${\bf b}$ the only change is $\omega=0.75 \omega_x\approx 1.06 \omega_y$ in the
second interval, such that the system enters there the unstable regime leading
to a  linear entanglement growth. This plot shows that entanglement can be
increased, kept constant and returned to a vanishing value just by tuning the
frequency $\omega$. } \label{f6}
\end{figure}

We finally show in Fig.\ \ref{f6} the possibilities offered by this model for
controlling the entanglement growth through a stepwise time dependent
frequency, starting from the separable ground state of $H_0$. After applying  a
``low'' initial frequency $\omega=0.5 \omega_x$ for $\omega_x t< 30$, which
leads to a weak quasiperiodic entanglement, by tuning $\omega$ to a value close
to the first instability $\omega_y=\sqrt{k_y}$ for  a finite time ($30<\omega_x
t<60$), it is possible to achieve a large entanglement increase (curve {\bf
a}). Then, by setting $\omega=0$ (i.e., switching off the field or rotation),
entanglement is kept high and constant, since the evolution operator becomes a
product of local mode evolutions. Finally, by turning the frequency on again up
to a low value, entanglement can be made to exhibit strong oscillations,
practically vanishing at the minimum if $\omega$ is appropriately tuned. Thus,
disentanglement at specific times can be achieved if desired. The entanglement
increase at the second interval can be enhanced by allowing the system to enter
the instability region for a short time, as shown in curve {\bf b}.

\vspace*{-0.cm}

\begin{figure}[t] \centerline{\hspace*{0.cm}\includegraphics[trim=0cm 0cm 0cm 0cm,
 width = 7cm]{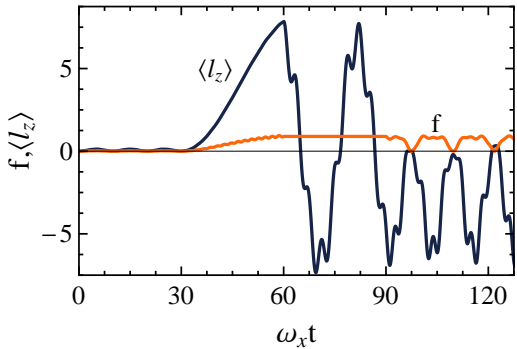}}\vspace*{0cm}
\caption{Evolution of the average occupation number $f(t)$ (Eq.\ (\ref{fmu}))
and angular momentum $\langle l_z\rangle$ for case {\bf a} of Fig.\ \ref{f6}.}
\label{f7}
\end{figure}
 \vspace*{-0.cm}

The growth of the average occupation $f(t)$ (and hence  the entanglement
entropy $S(t)$) in the second interval is strongly correlated with that of the
average angular momentum $\langle l_z\rangle_t$, i.e., with the entangling term
in $H$, as seen in Fig.\ \ref{f7} for case {\bf a} of Fig.\ \ref{f6}.
Nonetheless, while the evolution of $f(t)$ is similar to $S(t)$, the average
angular momentum exhibits pronounced oscillations  when $\omega$ is switched
off, since $l_z$ is not preserved in the present anisotropic trap
($[l_z,h_0]\neq 0$). These oscillations persist in the last interval, although
shifted and partly attenuated. Here  the vanishing of $\langle l_z\rangle_t$
provides a check for the vanishing entanglement, since in a separable state
$\langle l_z\rangle=\langle q_x\rangle\langle p_y\rangle-\langle q_y\rangle
\langle p_x\rangle$ and for the present initial conditions  $\langle
q_\mu\rangle=\langle p_\mu\rangle=0$ for all times. Thus, $f(t)=0$ implies here
$\langle l_z\rangle_t=0$, although the converse is not valid.

Though initially correlated, we remark that $\langle l_z\rangle_t$ and the
average occupation $f(t)$ do not have a fixed asymptotic relation in the whole
plane.  For instance, at the stability borders $\omega=\omega_\mu$, $\langle
l_z\rangle_t$ increases on average as $t^2$ for high $t$, i.e., as $f^2(t)$
(Eq.\ (\ref{flineal})), and the same relation with $f(t)$ holds in the unstable
sector {\bf D} ($\omega_y<\omega<\omega_x$), where $\langle l_z\rangle_t\propto
e^{2|\lambda_-|t}$. Nonetheless, in an unstable potential at the critical curve
$\Delta=0$,  $\langle l_z\rangle_t\propto t^2$ (on average) for large $t$,
increasing then as $f(t)$ (Eq.\ (\ref{f2t})), while at the critical points
${\bf L}$ we obtain $\langle l_z\rangle_t\propto t^6$, i.e., $\langle
l_z\rangle_t\propto f^{3/2}(t)$ asymptotically (Eq.\ (\ref{f4t})).

\section{Conclusions \label{IV}}
We have analyzed the entanglement generated by an angular momentum coupling
between two harmonic modes, when starting from a separable gaussian state. The
general treatment considered here is fully analytic and valid throughout the
entire parameter space, including stable and unstable regimes, as well as
critical regimes where the system cannot be written in terms of normal
coordinates or independent quadratic systems (non-diagonalizable ${\cal H}^2$).
Hence, in spite of its simplicity, the present model is able to exhibit
different types of entanglement evolution, including quasiperiodic evolution,
linear growth, and also logarithmic growth of the entanglement entropy with
time, which can all be reached just by tuning the frequency. The model is then
able to mimic the typical evolution regimes of the entanglement entropy
encountered in more complex many-body systems. Even distinct types of critical
logarithmic growth can be reached when allowing for general quadratic
potentials. The system offers then the possibility of an easily controllable
entanglement generation and growth, through stepwise frequency changes, which
can also be tuned in order to disentangle the system at specific times. The
model can therefore be of interest for continuous variable based quantum
information.

The authors acknowledge support from CONICET (LR,NC) and CIC (RR) of Argentina.
We also thank Prof.\ S.\ Mandal for motivating discussions during his visit to
our institute.

\subsection{Appendix}
In order to highlight the non-trivial character of the present model when
considered for all real values of the constants $k_\mu$ and frequency
$\omega>0$, we provide here some further details \cite{RK.09}. With the sole
exception of the critical curve $\Delta=0$ (Eq.\ (\ref{wc})), the Hamiltonian
(\ref{H1a}) can be written as a sum of two quadratic Hamiltonians,
 \begin{equation}
h=\frac{1}{2}(\alpha_{+}p_+^2+\beta_+
q_+^2)+\frac{1}{2}(\alpha_{-}p_-^2+\beta_- q_-^2) \,,\label{HD}\end{equation}
where $p_\pm$, $q_{\pm}$ are related with $q_{x,y}$, $p_{x,y}$ by the linear
canonical transformation
\begin{equation}
\begin{array}{rclrcl}
q_{\pm} &=&\frac{q_{x,y}-\eta p_{y,x}}{(1+\gamma\eta)}\,,&
 p_{\pm}&= & p_{x,y}+\gamma q_{y,x}\,, \label{ct}\\
\gamma&=&(\Delta-\varepsilon_-)/(2\omega)\,, &\eta&=&\gamma/\varepsilon_+\,,
\end{array}
\end{equation}
such that $[q_r,p_s]=i\delta_{rs}$, $[q_r,q_s]=[p_r,p_s]=0$ for
$r,s=\pm$, and
\[\alpha_{\pm}=\frac{1}{2}+\frac{\varepsilon_-\pm 2\omega^2}{2\Delta}\,,
 \,\,\,\,\,\,\,
 \beta_{\pm}=\frac{\Delta}{\omega^2}(\Delta\alpha_{\pm}-\varepsilon_-)\,,\]
with $\alpha_{\pm} \beta_{\pm}=\lambda_{\pm}^2$ (Eq.\ (\ref{la})). Nonetheless,
the coefficients  $\alpha_\pm$, $\beta_{\pm}$ can be positive, negative or
zero, and may become even complex, according to the values of $k_x$, $k_y$ and
$\omega$.  We may obviously interchange $p_\pm$ with $q_\pm$ in (\ref{ct}) by a
trivial canonical transformation $p_\pm\rightarrow q_\pm$, $q_\pm\rightarrow
-p_\pm$.  This freedom in the final form will be used in the following
discussion.

In sector {\bf A} of Fig.\ \ref{f1},  $\alpha_\pm$, $\beta_{\pm}$ in Eq.\
(\ref{HD}) are both {\it real and positive}, and the system is equivalent to
two harmonic modes. Here $\lambda_{\pm}$ are both real. In sector {\bf B},
$\alpha_+,\beta_+$ are positive but $\alpha_-$, $\beta_-$ are both negative,
so that the system is here equivalent to a standard plus an
``inverted'' oscillator. Nevertheless, $\lambda_{\pm}$ remain still real.
In sector {\bf C}, $\alpha_\pm\beta_{\pm}<0$, and the effective quadratic
potential becomes unstable in both coordinates (i.e., $\alpha_{\pm}>0$,
$\beta_{\pm}<0$). Here $\lambda_{\pm}$ are both imaginary. In sector {\bf D},
$\alpha_+$ and $\beta_+$ are positive but $\alpha_-\beta_-<0$,
so that the potential is stable in one direction but unstable in the other
(i.e., $\alpha_->0$, $\beta_-<0$).  Here $\lambda_+$ is real  but $\lambda_-$
is imaginary. In sector {\bf E}, both $\alpha_{\pm}$ and $\beta_{\pm}$ are
complex and $q_{\pm}$, $p_{\pm}$ are non hermitian \cite{RK.09}. Here
$\lambda_{\pm}$ are complex conjugates.

At the border {\bf A}-{\bf D}, $\alpha_+,\beta_+,\alpha_-$ are all positive but
$\beta_-=0$ (or similar with $\beta_-\leftrightarrow\alpha_-$), so that the
system is equivalent to a harmonic oscillator plus a free particle
($\lambda_+>0$, $\lambda_-=0$). The same holds at the border {\bf B}-{\bf D},
except that $\alpha_-<0$ (inverted free particle term).  At the Landau point
{\bf F}  ($k_x=k_y=\omega^2$, where {\bf A}, {\bf B}, and {\bf D} meet),
$\alpha_+>0$, $\beta_+>0$ but $\alpha_-=\beta_-=0$. Finally, at the border {\bf
C}-{\bf D}, $\alpha_+>0$, $\beta_+=0$ (or similar with
$\beta_+\leftrightarrow\alpha_+$) and  $\alpha_-\beta_-<0$, implying
$\lambda_+=0$ and $\lambda_-$ imaginary.

The decomposition (\ref{HD}) no longer holds at the critical curve $\Delta=0$,
which separates sector {\bf E} from sectors {\bf C} and {\bf D}. At this curve
(including points {\bf L}), the system is {\it inseparable}, in the sense that
it cannot be written as a sum of two independent quadratic systems, even if
allowing for complex coordinates and momenta as in sector {\bf E}. While the
matrix  ${\cal H}^2$ (Eq.\ (\ref{H2})) is always diagonalizable for $\Delta\neq
0$, i.e., whenever the decomposition (\ref{HD}) is feasible, both ${\cal
H}$ {\it and} ${\cal H}^2$ are non-diagonalizable when $\Delta=0$. Here
$\lambda_\pm=\lambda$, with $\lambda$ real at the border between {\bf B} and
{\bf E}, imaginary between {\bf C} and {\bf E} and zero at the points {\bf L}.
The optimum decomposition of $h$ in these cases is discussed in \cite{RK.09}.


\begin{thebibliography}{99}
\bibitem{SLRD.13}J.\ Schachenmayer, B.P.\ Lanyon,  C.F.\ Roos,  A.J.\ Daley,
Phys.\ Rev.\ X {\bf 3} 031015 (2013).
\bibitem{BPM.12} J.H.\ Bardarson, F.\ Pollmann, J.E.\ Moore,
Phys.\ Rev.\ Lett.\ {\bf 109} 017202 (2012).
\bibitem{DPSZ.12} A.J.\ Daley, H.\ Pichler, J.\ Schachenmayer, P. Zoller,
Phys.\ Rev.\ Lett.\ {\bf 109} 020505 (2012).
\bibitem{Be.93} C.H.\ Bennett et al., Phys.\ Rev.\ Lett.\ {\bf 70}, 1895
    (1993); Phys.\ Rev.\ Lett.\ {\bf 76}, 722 (1996).
\bibitem{NC.00}M.A.\ Nielsen and I.L.\ Chuang, {\it Quantum Computation and
        Quantum Information} (Cambridge Univ.\ Press, Cambridge, UK, 2000).
\bibitem{JL.03} R.\ Josza and N.\ Linden, Proc.\ R.\ Soc.\ {\bf A 459},
2011 (2003).
\bibitem{Vi.03} G.\ Vidal, Phys.\ Rev.\ Lett.\ {\bf 91}, 147902 (2003).
\bibitem{SWVC.08}N. Schuch, M. M.Wolf, F. Verstraete, and J. I. Cirac, {\it Phys.\
Rev.\  Lett.\ } {\bf 100}, 030504 (2008).
\bibitem{Va.56} J.G. Valatin, Proc.\ R.\ Soc.\ London {\bf 238}, 132 (1956).
\bibitem{FK.70} A.\ Feldman and A.\ H.\ Kahn, Phys.\ Rev.\ B {\bf 1}, 4584 (1970).
\bibitem{RS.80} P.~Ring and P.~Schuck, {\it The Nuclear Many-Body Problem},
(Springer, NY, 1980).
\bibitem{BR.86} J.P. Blaizot and G. Ripka, {\it Quantum Theory of Finite
Systems} (MIT Press, MA, 1986).
\bibitem{MC.94}A.V.\ Madhav, T.\ Chakraborty, Phys.\ Rev.\ B {\bf 49}, 8163 (1994).
\bibitem{LNF.01} M.\ Linn, M.\ Niemeyer, and A.\ L.\ Fetter,
Phys.\ Rev.\ A {\bf 64}, 023602 (2001).
\bibitem{OO.04} M.\ \"O.\ Oktel, Phys.\ Rev.\ A {\bf 69}, 023618 (2004).
\bibitem{AF.07} A.\ L.\ Fetter, Phys.\ Rev.\ A {\bf 75}, 013620 (2007).
\bibitem{ABL.09} A.\ Aftalion, X.\ Blanc, and N.\ Lerner,
Phys.\ Rev.\ A {\bf 79}, 011603(R) (2009).
\bibitem{ABD.05} A.\ Aftalion, X.\ Blanc, J.\ Dalibard,
Phys.\ Rev.\ A {\bf 71}, 023611 (2005);
S. Stock et al, Laser Phys.\ Lett. {\bf 2}, 275 (2005).
\bibitem{BDS.08}
A.\L.\ Fetter, Rev.\ Mod.\ Phys.\ {\bf 81}, 647 (2009).
I.\ Bloch, J.\ Dalibard, W.\ Zwerger, Rev.\ Mod.\ Phys.\ {\bf 80}, 885 (2008);
\bibitem{RK.09} R.\ Rossignoli and A.M.\ Kowalski, Phys.\ Rev.\ A {\bf 79}
062103 (2009).
\bibitem{WW.01} R.F.\ Werner, M.M.\ Wolf, Phys.\ Rev.\ Lett.\ {\bf 86}, 3658 (2001).
 \bibitem{AEPW.02} K.\ Audenaert, J.\ Eisert, M.B.\ Plenio, and R.F.\ Werner,
 Phys.\ Rev.\ A {\bf 66}, 042327 (2002).
 \bibitem{ASI.04}G.~Adesso, A.~Serafini, and F.~Illuminati,
Phys.~Rev.~A {\bf 70}, 022318 (2004); A.~Serafini, G.~Adesso, and
F.~Illuminati, Phys.~Rev.~A {\bf 71}, 032349 (2005).
\bibitem{BvL.05}S.L.\ Braunstein and P.\ van Loock,
 Rev.\ Mod.\ Phys.\ {\bf 77}, 513 (2005).
 \bibitem{WPPCRSL.12} C.\ Weedbrook et al, Rev.\ Mod.\ Phys.\ {\bf 84}, 621.
 \bibitem{PE.94}J.\ P\v{e}rina, Z.\ Hradil, and B.\ Jur\v{c}o,
{\it Quantum optics and Fundamentals of Physics} (Kluwer, Dordrecht, 1994);
N.\ Korolkova, J.\ P\v{e}rina, Opt.\ Comm.\ {\bf 136}, 135 (1996).
\bibitem{LR.11}L.\ Reb\'on, R. Rossignoli, Phys.\ Rev.\ A {\bf 84}, 052320 (2011).
\bibitem{HMM.03} A.P.\ Hines, R.\ H.\ McKenzie, and G.J.\ Milburn,
Phys.\ Rev.\ A {\bf 67}, 013609 (2003).
\bibitem{NL.05}H.T.\ Ng, P.T.\ Leung, Phys.\ Rev.\ A {\bf 71}, 013601 (2005).
\bibitem{CN.08} A.V.\ Chizhov, R.G.\ Nazmitdinov, Phys.\ Rev.\ A {\bf 78},
064302 (2008).
\bibitem{SGLK.12} V.\ Sudhir, M.G.\ Genoni, J.\ Lee, M.S.\ Kim,
Phys.\ Rev.\ A {\bf 86}, 012316 (2012).
\bibitem{RS.07} R. Rossignoli and C.T. Schmiegelow, Phys.\ Rev.\ A {\bf 75}, 012320 (2007).
\bibitem{AOPFP.04} L. Amico, et al, Phys.\ Rev.\ A {bf 69}, 022304 (2004);
 A. Sen(De), U. Sen, and M. Lewenstein, Phys.\ Rev.\ A {\bf 70},
060304(R) (2004); {\bf 72}, 052319 (2005).
\bibitem{HK.05}S.D.\ Hamieh and M.I.\ Katsnelson, Phys.\ Rev.\ A {\bf 72}, 032316
(2005); Z. Huang, S. Kais, Phys.\ Rev.\ A {\bf 73}, 022339 (2006);
 M.\ Koniorczyk, P.\ Rapcan,  V.\ Buzek, Phys.\ Rev.\ {\bf A} 72,
022321 (2005).
\bibitem{AFOV.08} L.\ Amico, R.\ Fazio, A.\ Osterloh and
           V.\ Vedral, Rev.\ Mod.\ Phys.\ {\bf  80}, 516  (2008).
\bibitem{EA.02} A.K.\ Ekert et al, Phys.\ Rev.\ Let..\ {\bf 88}, 217901 (2002);
C.M.\ Alves, D.\ Jaksch, Phys.\ Rev.\ Let..\ {\bf 93}, 110501 (2004).
    \end{thebibliography}
\end{document}